\documentclass{ws-procs975x65}            % default, citations in superscript
\begin{document}

\makeatletter
\newcommand{\figcaption}{\def\@captype{figure}\caption}
\newcommand{\tabcaption}{\def\@captype{table}\caption}
\newcommand{\rmnum}[1]{\romannumeral #1}
\newcommand{\Rmnum}[1]{\expandafter\@slowromancap\romannumeral #1@}
\def\hlinewd#1{%
  \noalign{\ifnum0=`}\fi\hrule \@height #1 \futurelet
   \reserved@a\@xhline}
\makeatother

\def\dab{\int^{\alpha_{max}}_{\alpha_{min}}d\alpha\int^{\beta_{max}}_{\beta_{min}}d\beta}
\def\qq{\langle\bar qq\rangle}
\def\GGa{\langle GG\rangle}
\def\GGb{\langle g_s^2GG\rangle}
\def\GGGa{\langle GGG\rangle}
\def\GGGb{\langle g_s^3fGGG\rangle}
\def\qGqa{\langle\bar  qGq\rangle}
\def\qGqb{\langle\bar qg_s\sigma\cdot Gq\rangle}
\def\JJa{\langle jj\rangle}
\def\JJb{\langle g_s^4jj\rangle}
\def\efun{e^{-\frac{m_c^2}{\alpha(1-\alpha)M_B^2}}}
\def\f(s){\left[(\alpha+\beta)m_c^2-\alpha\beta s\right]}
\def\non{\\ \nonumber}

\title{Exotic hadron states}

\author{Wei Chen$^*$, J. Ho and T. G. Steele}
\address{Department of Physics
and Engineering Physics, University of Saskatchewan, Saskatoon,\\
Saskatoon, SK, S7N 5E2, Canada\\
$^*$E-mail: wec053@mail.usask.ca}

\author{R. T. Kleiv, B. Bulthuis, D. Harnett and T. Richards}
\address{Department of Physics, University of the Fraser Valley,\\
Abbotsford, BC, V2S 7M8, Canada}

\author{Shi-Lin Zhu}
\address{School of Physics and State Key Laboratory of Nuclear Physics and Technology, Peking University, \\
Beijing 100871, China\\
Collaborative Innovation Center of Quantum Matter, \\
Beijing 100871, China\\
Center of High Energy Physics, Peking University,\\
Beijing 100871, China}

\begin{abstract}
Many charmonium-like and bottomonium-like $XYZ$ resonances have been
observed by the Belle, Babar, CLEO and BESIII collaborations in the
past decade. They are difficult to fit in the conventional quark
model and thus are considered as candidates of exotic hadrons, such
as multi-quark states, meson molecules, and hybrids. In this talk,
we first briefly introduce the method of QCD sum rules and then
provide a short review of the mass spectra of the quarkonium-like
tetraquark states and the heavy quarkonium hybrids in the QCD sum
rules approach. Possible interpretations of the $XYZ$ resonances are
briefly discussed.
\end{abstract}


\bodymatter

\section{Introduction}\label{sec1}
In the conventional quark model, mesons are composed of a pair of
quark and antiquark ($q\bar q$) while baryons are composed of three
quarks ($qqq$). Hadrons with different quark contents from $q\bar q$
or $qqq$ are called exotic states, such as glueballs, hybrids,
multiquarks and so on. Although these exotic hadron configurations
are allowed in QCD, there was no significant evidence of their existence 
until recently. 

Since 2003, many new charmonium-like and bottomonium-like states
were observed experimentally (see
Refs.\cite{2014-Chen-p13-40,2013-Bodwin-p-,2012-Faccini-p1230025-1230025}
for recent reviews). These new states were observed through either
the hidden-charm/hidden-bottom or open-charm/open-bottom final
states and thus contain a heavy quark-antiquark pair. However, some
of these states do not fit in the conventional quark model and are
considered as candidates for exotic states. To explore the
underlying structures of these XYZ states, many theoretical
speculations have been proposed such as the molecular states,
quarkonium-like tetraquark states, quarkonium hybrids and
conventional quarkonium states.

Hadronic molecules are loosely bound states composed of two heavy
mesons$(Q\bar q)(\bar Qq)$, in which $Q$ represents a heavy quark
(charm or bottom quark) and $q$ a light quark (up, down or strange
quark). They are probably bound by the long-range color-singlet pion
exchange. The masses of the molecules are slightly lower than some
open-flavor thresholds. Some XYZ states, such as $X(3872)$ and the
charged $Z_c, Z_b$ states, were discovered to be very close to the
open-charm/bottom thresholds. They were naturally considered as
candidates of the molecular states. Tetraquarks are composed of a
pair of diquark and antidiquark $(Qq)(\bar Q\bar q)$, which are
bound by the colored force between quarks. They can decay into a
pair of heavy mesons or one charmonium/bottomonium plus a light
meson through rearrangement process. Thus the tetraquarks are
expected to be very broad resonances \cite{2013-Chen-p-}. Hybrids
are bound states of a quark-antiquark pair and an excited gluon.
The $Y(4260)$ meson was considered as a compelling candidate of the
charmonium hybrid
\cite{2005-Zhu-p212-212,2005-Kou-p164-169,2005-Close-p215-222}. The
spectroscopy of these exotic hadrons should be studied
systematically to improve our understanding of the relations between
the exotic hadrons and the newly observed XYZ states.

In this talk, we focus on the quarkonium-like tetraquark states and the quarkonium hybrid states.
We briefly introduce our recent results of the mass spectra of these systems. We then try to understand
the nature of some XYZ states using the formalisms of tetraquark states and charmonium hybrids.

\section{QCD sum rules}
In the past several decades, QCD sum rules have been proven to be a very powerful non-perturbative tool
to study the hadron structures\cite{1979-Shifman-p385-447,1985-Reinders-p1-1,2000-Colangelo-p1495-1576,
2014-Chen-p13-40}. Considering the two-point correlation function induced by two
hadronic currents
\begin{eqnarray}
\Pi(q^{2})= i\int d^4xe^{iq\cdot x}\langle0|T[J(x)J^{\dag}(0)]|0\rangle, \label{equ:Pi}
\end{eqnarray}
in which $J(x)$ is an interpolating current carrying the same
quantum numbers as the hadrons we want to study. The basic idea of
QCD sum rule approach is that this two-point function can be
achieved at both the hadron level and the quark-gluon level. A
fundamental assumption is the quark-hadron duality, which ensures
the equivalence of the correlation functions obtained at these two
levels.

At the hadron level, the correlation function is described via the dispersion relation
\begin{eqnarray}
\Pi(q^2)=(q^2)^N\int_{4m_c^2}^{\infty}\frac{\rho(s)}{s^N(s-q^2-i\epsilon)}ds+\sum_{n=0}^{N-1}b_n(q^2)^n, \label{dispersion_relation}
\end{eqnarray}
where $b_n$ is the unknown subtraction constant which can be removed by taking the Borel transform.
$\rho(s)$ is the spectral function in the narrow resonance approximation
\begin{eqnarray}
\nonumber
\rho(s)&\equiv&\sum_n\delta(s-m_n^2)\langle0|J|n\rangle\langle n|J^{\dagger}|0\rangle
\\&=&f_X^2\delta(s-m_X^2)+ \mbox{continuum}.  \label{Phenrho}
\end{eqnarray}

The two-point function $\Pi(q^2)$ can also be calculated at the quark-gluonic level via the operator
product expansion (OPE)
\begin{equation}
\Pi(q)= i \int d^4xe^{iq\cdot x}\langle
0|T[J(x)J^{\dag}(0)]|0\rangle=\sum_nC_n(Q^2)O_n\,,~Q^2=-q^2\,,
\end{equation}
in which $C_n(Q^2)$ are the Wilson coefficients and $O_n$ are the various QCD condensates,
including the quark condensate $\qq$, gluon condensate $\GGb$, quark-gluonmixed condensate
$\qGqb$, tri-gluon condensate $\GGGb$, four quark condensate $\qq^2$, and dimension-8 condensate
$\qq\qGqb$. In general, the Wilson coefficients $C_n(Q^2)$ can be calculated in perturbation theory
and expressed in terms of the QCD parameters such as the quark mass and the strong coupling constant
$\alpha_s$. The long distance non-perturbative effects are included in the various condensates $O_n$,
which are ordered by increasing dimension in the expansion.

We then can obtain information on hadron properties by equating the two-point correlation functions at 
these two levels. The mass sum rules are usually established after performing Borel transform of both sides
\begin{eqnarray}
f_X^2m_X^{2k}e^{-m_X^2/M_B^2}=\int_{4m_Q^2}^{s_0}dse^{-s/M_B^2}\rho(s)s^k=\mathcal{L}_{k}\left(s_0,
M_B^2\right), \label{sumrule}
\end{eqnarray}
where $s_0, M_B$ are the continuum threshold and Borel mass respectively and $\rho(s)=\frac{1}{\pi}\text{Im}\Pi(s)$.
The lowest lying hadron mass is then extracted as
\begin{eqnarray}
m_X\left(s_0, M_B^2\right)=\sqrt{\frac{\mathcal{L}_{1}\left(s_0,
M_B^2\right)}{\mathcal{L}_{0}\left(s_0, M_B^2\right)}}. \label{mass}
\end{eqnarray}

In the next two sections, we will introduce the QCD sum rule study
of the mass spectra of the quarkonium-like tetraquark states and
quarkonium hybrid states.

\section{Quarkonium-like tetraquark states}
The diquark ($qq$) concept plays a very important role in the
study of the tetraquark states. They are the bricks used to construct a
tetraquark field ($qq\bar q\bar q$). The properties of diquark
fields, including their spins, parities, flavor, color and Lorentz
structures were studied in
Refs.\cite{2005-Jaffe-p1-45,2013-Du-p14003-14003} Tetraquarks
($qq\bar q\bar q$) are then composed of diquarks and antidiquarks.
The low-lying scalar mesons below 1 GeV have been considered as good
candidates of the light tetraquark
states\cite{2007-Chen-p369-372,2007-Chen-p94025-94025}. In the heavy
quark sector, some of the recently observed quarkonium-like states
were suggested to be candidates of hidden charm/bottom $Q{\bar
Q}q\bar{q}$-type tetraquark states \cite{2007-Matheus-p14005-14005,
2007-Maiani-p182003-182003,2006-Ebert-p214-219,2010-Chen-p105018-105018,
2011-Chen-p34010-34010,2013-Du-p33104-33104,Kleiv:2013dta}. 

\subsection{$X(3872)$ and the charged $Z_c, Z_b$ states}
$X(3872)$ was the first charmonium-like state discovered in B-factories. It was reported in 2003 by Belle Collaboration
in the $J/\psi\pi^+\pi^-$ final states on the process of $B^+\to K^+J/\psi\pi^+\pi^-$\cite{2003-Choi-p262001-262001}.
Ten years after that, the LHCb Collaboration determined its quantum numbers as $J^{PC}=1^{++}$\cite{2013-Aaij-p222001-222001}.

Recently, the family of the charged quarkonium-like states has become more abundant.
The first charged state $Z(4430)^+$ was observed in the $\psi(2S)\pi^+$ invariant mass spectrum in the
process $\bar B^0\to \psi(2S)\pi^+K^-$ by the Belle Collaboration~\cite{2008-Choi-p142001-142001} and confirmed
recently by the LHCb Collaboration~\cite{2014-Aaij-p222002-222002}. Later,
%Belle Collaboration reported $Z(4050)^+$ and $Z(4250)^+$\cite{2008-Mizuk-p72004-72004} and
The BESIII Collaboration reported
$Z_c(3900)^+$\cite{2013-Ablikim-p252001-252001}, which was confirmed
quickly by the Belle Collaboration\cite{2013-Liu-p252002-252002} and by
using CLEO data\cite{2013-Xiao-p366-370}. The BESIII Collaboration also
observed $Z_c(4025)^{\pm}$\cite{2014-Ablikim-p132001-132001} and
$Z_c(4020)^{\pm}$ \cite{2013-Ablikim-p242001-242001} later. This
year, the Belle Collaboration reported two new charged states
$Z_c(4200)^+$ \cite{2014-Chilikin-p-} and
$Z_c(4050)^+$\cite{2014-Wang-p-}. Moreover, the Belle Collaboration also
observed two charged bottomonium-like states $Z_b(10610)$ and
$Z_b(10650)$\cite{2011-Adachi-p-}. All these resonances have the
exotic flavor content $c\bar cu\bar d$ for $Z_c$ states and $b\bar
bu\bar d$ for $Z_b$ states. They are isovector axial-vector states
with positive $G$-parities. Thus their neutral partners carry the
quantum numbers $I^GJ^{PC}=1^+1^{+-}$.

%%%%%%%%%%%%%%%%%%%%%%%%%%%%%%%%%%%%%%%%%%%%%%%%%%%%%%%%%%%%%%%%
\begin{table}
\tbl{Mass spectra for the $qc\bar q\bar c$ and $qb\bar q\bar b$ tetraquark states with $J^{PC}=1^{++}$.}
%\begin{center}
%\renewcommand{\arraystretch}{1.3}
%\begin{tabular*}{13cm}{cccccc}
%\hlinewd{.8pt}
{\begin{tabular}{@{}cccccc@{}}
\toprule
                   & Currents & $s_0(\mbox{GeV}^2)$&$[M^2_{\mbox{min}}$,$M^2_{\mbox{max}}](\mbox{GeV}^2)$&$m_X$\mbox{(GeV)}&PC(\%)\\
\colrule
$qc\bar q\bar c$ system & $J_{3\mu}$         &  $4.6^2$         & $3.0-3.4$           & $4.19\pm0.10$     & 47.3 \\
                        & $J_{4\mu}$         &  $4.5^2$         & $3.0-3.3$           & $4.03\pm0.11$     & 46.8
\vspace{5pt} \\
$qb\bar q\bar b$ system & $J_{4\mu}$         &  $10.8^2$        & $8.5-9.2$           & $10.22\pm0.11$    & 44.6 \\
                        & $J_{7\mu}$         &  $10.7^2$        & $7.8-8.4$           & $10.14\pm0.10$    & 44.8  \\
                        & $J_{8\mu}$         &  $10.7^2$        & $7.8-8.4$           & $10.14\pm0.09$    & 44.8  \\
%\hlinewd{.8pt}
%\end{tabular*}
\botrule
\end{tabular}
} \label{table1++}
%\end{center}
\end{table}
%%%%%%%%%%%%%%%%%%%%%%%%%%%%%%%%%%%%%%%%%%%%%%%%%%%%%%%%%%%%%%%%
%%%%%%%%%%%%%%%%%%%%%%%%%%%%%%%%%%%%%%%%%%%%%%%%%%%%%%%%%%%%%%%%
\begin{table}
\tbl{Mass spectra for the $qc\bar q\bar c$ and $qb\bar q\bar b$ tetraquark states with $J^{PC}=1^{+-}$.}
%\begin{center}
%\renewcommand{\arraystretch}{1.3}
%\begin{tabular*}{13cm}{cccccc}
%\hlinewd{.8pt}
{\begin{tabular}{@{}cccccc@{}}
\toprule
                   & Currents & $s_0(\mbox{GeV}^2)$&$[M^2_{\mbox{min}}$,$M^2_{\mbox{max}}](\mbox{GeV}^2)$&$m_X$\mbox{(GeV)}&PC(\%)\\
\colrule
                        & $J_{3\mu}$         &  $4.6^2$            & $3.0-3.4$           & $4.16\pm0.10$     & 46.2  \\
$qc\bar q\bar c$ system & $J_{4\mu}$         &  $4.5^2$            & $3.0-3.3$           & $4.02\pm0.09$     & 44.6  \\
                        & $J_{5\mu}$         &  $4.5^2$            & $3.0-3.4$           & $4.00\pm0.11$     & 46.0  \\
                        & $J_{6\mu}$         &  $4.6^2$            & $3.0-3.4$           & $4.14\pm0.09$     & 47.0
\vspace{5pt} \\
                        & $J_{3\mu}$         &  $10.6^2$           & $7.5-8.5$           & $10.08\pm0.10$    & 45.9  \\
$qb\bar q\bar b$ system & $J_{4\mu}$         &  $10.6^2$           & $7.5-8.5$           & $10.07\pm0.10$    & 46.2  \\
                        & $J_{5\mu}$         &  $10.6^2$           & $7.5-8.4$           & $10.05\pm0.10$    & 45.3  \\
                        & $J_{6\mu}$         &  $10.7^2$           & $7.5-8.7$           & $10.15\pm0.10$    & 47.6  \\
%\hlinewd{.8pt}
%\end{tabular*}
\botrule
\end{tabular}
} \label{table1+-}
%\end{center}
\end{table}
%%%%%%%%%%%%%%%%%%%%%%%%%%%%%%%%%%%%%%%%%%%%%%%%%%%%%%%%%%%%%%%%%%%%%%%
Could these charged $Z_c, Z_b$ states and $X(3872)$ be
quarkonium-like tetraquark states? In
Refs.\cite{2011-Chen-p34010-34010,2012-Chen-p1003-1003}, we
constructed all the tetraquark interpolating currents with
$J^{PC}=1^{++}, 1^{+-}$ in a systematic way. Then we calculated
the two-point correlation functions and performed QCD sum rules
analyses of these systems. The mass spectra of the charmonium-like
and bottomonium-like tetraquark states with $J^{PC}=1^{++}$ are
given in Table \ref{table1++}. For the current $J_{4\mu}$, the mass
of the $qc\bar q\bar c$ state was extracted as $m_X=4.03\pm0.11$
GeV, which is slightly higher than the mass of $X(3872)$. In Table
\ref{table1+-}, we give the mass spectra for the $qc\bar q\bar c$
and $qb\bar q\bar b$ tetraquark states with $J^{PC}=1^{+-}$. It was
shown that the masses for the charmonium-like tetraquark states were
$m_X=3.9-4.3$ GeV. These masses are consistent with the
mass regions of the charged $Z_c$ states. In other words, our result
supports these charged resonances, such as $Z_c(3900), Z_c(4025),
Z_c(4050)$ and $Z_c(4200)$, to be the candidates of the
charmonium-like tetraquark states. However, it is difficult to
differentiate these charged states in the two-point sum rule
analysis. Instead, one needs to calculate the three-point functions
to study the hadronic three-point interaction vertices
\cite{2014-chen-}. The coupling constants and hadronic decay widths
of these states can then be determined.

For the bottomonium-like $qb\bar q\bar b$ systems with
$J^{PC}=1^{++}$ and $1^{+-}$, their masses were both extracted as
$m_{X^b}=10.0-10.3$ GeV, which were much lower than the masses of
$Z_b(10610)$ and $Z_b(10650)$.

The numerical results for the other channels can be found in
Refs.\cite{2010-Chen-p105018-105018,2011-Chen-p34010-34010,2013-Du-p33104-33104,2013-Du-p14003-14003,2014-Chen-p54037-54037}.

\section{Charmonium and bottomonium hybrid states}
The heavy quarkonium hybrids were firstly studied in
Refs.~\cite{1985-Govaerts-p215-215,1985-Govaerts-p575-575,
1987-Govaerts-p674-674} by using QCD sum rule approach, in which
only the leading-order contributions of the perturbative terms and
the dimension four gluonic condensate were considered. The mass sum
rules of some $J^{PC}$ channels were unstable and thus these mass
predictions were unreliable. Recently, the
$J^{PC}=1^{--}$\cite{2012-Qiao-p15005-15005},
$1^{++}$\cite{2012-Harnett-p125003-125003} and
$0^{-+}$\cite{2012-Berg-p34002-34002} channels have been re-analyzed
by including the dimension six tri-gluon condensate. The dimension
six contributions have proven to be very important because they
stabilize the hybrid sum rules. In Ref.\cite{2013-Chen-p19-19}, we
have extended the calculations to the remaining channels. After
performing the QCD sum-rule analysis, we updated the mass spectra of
charmonium and bottomonium hybrids with exotic and non-exotic
quantum numbers in Table \ref{tablecchybrid} and Table
\ref{tablebbhybrid}, respectively.
%%%%%%%%%%%%%%%%%%%%%%%%%%%%%%%%%%%%%%%%%%%%%%%%%%%%%%%%%%%%%%%%%%%%
\begin{table}
\tbl{Mass spectrum of the charmonium hybrid states.}
%\begin{center}
%\renewcommand{\arraystretch}{1.3}
%\begin{tabular*}{13cm}{cccccc}
%\hlinewd{.8pt}
{\begin{tabular}{@{}cccccc@{}}
\toprule
& ~~$J^{PC}$~~ & $s_0(\mbox{GeV}^2)$&$[M^2_{\mbox{min}}$,$M^2_{\mbox{max}}](\mbox{GeV}^2)$&$m_X$\mbox{(GeV)}&~~~~PC(\%)\\
\colrule
& $1^{--}$   & 15  & $2.5-4.8 $& $3.36\pm0.15$& 18.3\\
& $0^{-+}$   & 16  & $5.6-7.0 $& $3.61\pm0.21$& 15.4\\
& $1^{-+}$   & 17  & $4.6-6.5 $& $3.70\pm0.21$& 18.8\\
& $2^{-+}$   & 18  & $3.9-7.2 $& $4.04\pm0.23$& 26.0
\vspace{5pt}\\
& $0^{+-}$   & 20  & $6.0-7.4 $& $4.09\pm0.23$& 15.5\\
& $2^{++}$   & 23  & $3.9-7.5$ & $4.45\pm0.27$& 21.5\\
& $1^{+-}$   & 24  & $2.5-8.4 $& $4.53\pm0.23$& 33.2\\
& $1^{++}$   & 30  & $4.6-11.4$& $5.06\pm0.44$& 30.4\\
& $0^{++}$   & 34  & $5.6-14.6$& $5.34\pm0.45$& 36.3
\vspace{5pt}\\
& $0^{--}$    & 35  & $6.0-12.3$& $5.51\pm0.50$& 31.0\\
%\hlinewd{.8pt}
%\end{tabular*}
\botrule
\end{tabular}
} \label{tablecchybrid}
%\end{center}
\end{table}
%%%%%%%%%%%%%%%%%%%%%%%%%%%%%%%%%%%%%%%%%%%%%%%%%%%%%%%%%%%%%%%

It was shown that the negative-parity states with $J^{PC}=(0, 1,
2)^{-+}, 1^{--}$ form the lightest hybrid supermultiplet while the
positive-parity states with $J^{PC}=(0, 1)^{+-}, (0, 1, 2)^{++}$
belong to a heavier hybrid supermultiplet, confirming the
supermultiplet structure found in the other
approaches\cite{2012-Liu-p126-126}. The hybrid with $J^{PC}=0^{--}$
has a much higher mass which may suggest a different excitation of
the gluonic field compared to the other channels. Consistent with
the previous results, we found that the $J^{PC}=1^{++}$ charmonium
hybrid is substantially heavier than the $X(3872)$, which seems to
preclude a pure charmonium hybrid interpretation for this state. The
open-flavour bottom-charm $\bar bGc$ hybrid systems were also
studied in Ref.\cite{2014-Chen-p25003-25003}.
%%%%%%%%%%%%%%%%%%%%%%%%%%%%%%%%%%%%%%%%%%%%%%%%%%%%%%%%%%%%%%%
\begin{table}
\tbl{Mass spectrum of the bottomonium hybrid states.}
%\begin{center}
%\renewcommand{\arraystretch}{1.3}
%\begin{tabular*}{13cm}{cccccc}
%\hlinewd{.8pt}
{\begin{tabular}{@{}cccccc@{}}
\toprule
& ~~$J^{PC}$ ~~& $s_0(\mbox{GeV}^2)$&$[M^2_{\mbox{min}}$,$M^2_{\mbox{max}}](\mbox{GeV}^2)$&$m_X$\mbox{(GeV)}&~~~~PC(\%)\\
\colrule
& $1^{--}$    & 105  & $11-17 $& $9.70\pm0.12$&  17.2\\
& $0^{-+}$   & 104  & $14-16 $& $9.68\pm0.29$& 17.3\\
& $1^{-+}$   & 107  & $13-19 $& $9.79\pm0.22$& 20.4\\
& $2^{-+}$   & 105  & $12-19$& $9.93\pm0.21$& 21.7
\vspace{5pt}\\
& $0^{+-}$   & 114  & $14-19 $& $10.17\pm0.22$& 17.6\\
& $2^{++}$   & 120  & $12-20$ & $10.64\pm0.33$& 19.7\\
& $1^{+-}$   & 123  & $10-21 $& $10.70\pm0.53$& 28.5\\
& $1^{++}$  & 134  & $13-27$& $11.09\pm0.60$& 27.7\\
& $0^{++}$  & 137  & $13-31$& $11.20\pm0.48$& 30.0
\vspace{5pt}\\
& $0^{--}$    & 142  & $14-25$& $11.48\pm0.75$& 24.1\\
%\hlinewd{.8pt}
%\end{tabular*}
\botrule
\end{tabular}
} \label{tablebbhybrid}
%\end{center}
\end{table}
%%%%%%%%%%%%%%%%%%%%%%%%%%%%%%%%%%%%%%%%%%%%%%%%%%%%%%%%%%%%%%%%%%%

\section*{Acknowledgments}
This project was supported by the Natural Sciences and Engineering Research Council of Canada (NSERC)
and the National Natural Science Foundation of China under Grant NO. 11261130311.

%\bibliographystyle{ws-procs975x65}
%\bibliography{myreference}

%Non BiBTeX users can list down their references as:

\end{document}